\documentclass[]{interact}

\usepackage{epstopdf}

\usepackage[numbers,sort&compress]{natbib}
\bibpunct[, ]{[}{]}{,}{n}{,}{,}
\renewcommand\bibfont{\fontsize{10}{12}\selectfont}
\makeatletter
\def\NAT@def@citea{\def\@citea{\NAT@separator}}
\makeatother

\theoremstyle{plain}
\newtheorem{theorem}{Theorem}[section]
\newtheorem{lemma}[theorem]{Lemma}

\newtheorem{proposition}[theorem]{Proposition}
\newtheorem{assum}{Assumption}

\theoremstyle{definition}
\newtheorem{definition}[theorem]{Definition}

\graphicspath{{figures/}}

\theoremstyle{remark}

\usepackage{caption}
\usepackage{subcaption}
\usepackage{graphicx}
\usepackage{algorithm}
\usepackage{algorithmic}
\usepackage{tikz}
\usetikzlibrary{arrows, automata}
\usepackage{titlecaps}
\usepackage{multirow}

\DeclareMathOperator*{\argmin}{arg\,min}

\newcommand\Pro{\mathbf{P}}
\newcommand\Gi{\mathcal{G}}
\newcommand\X{\mathbf{X}}
\newcommand\Xt{\X(t)}
\newcommand\Xto{\X(t-1)}
\newcommand\Xtp{\X(t-p_0)}

\newcommand\ut{\mathbf{u}(t)}
\newcommand\varep{\boldsymbol{\varepsilon}}
\newcommand\bol{\boldsymbol}

\begin{document}

\articletype{}

\title{On Learning Time Series Summary DAGs: A Frequency Domain Approach}

\author{
\name{Aramayis Dallakyan\textsuperscript{a}\thanks{CONTACT Aramayis DAllakyan. Email: dallakyan.aramayis@gmail.com}}
\affil{\textsuperscript{a} StataCorp, College Station, Texas, USA}
}

\maketitle

\begin{abstract}
The fields of time series and graphical models emerged and advanced separately. Previous work on the structure learning of
 continuous and real-valued time series utilizes the time domain, with a focus on either structural autoregressive models or linear
  (non-)Gaussian Bayesian Networks. In contrast,  we propose a novel frequency domain approach to identify a
   topological ordering and learn the structure of both real and complex-valued multivariate time series. 
   In particular, we define a class of complex-valued  Structural Causal Models (cSCM)
    at each frequency of the Fourier transform of the time series.
Assuming that the time series is generated from the transfer function model, we show that the topological ordering and
 corresponding summary directed acyclic graph can be uniquely identified from cSCM. 
 The performance of our algorithm
  is investigated using simulation experiments and real datasets. Code implementing the proposed algorithm is available at Supplementary Materials.
\end{abstract}


\section{Introduction}

Structure learning in time series is used in many applications such as machine learning \citep{peters2017}, economics \citep{bessler2003,  demiralp2003}, 
climate research \citep{runge2019b}, and earth science \citep{runge2019a}.  There are two general approaches depending on the time-resolution of the data
\citep{breitung2002, rajaguru2008,hyvarinen2010}. First, if the time-resolution of the measurements is higher than the time scale of the causal influence,  
then the structure can be learned from the autoregressive model with time-lagged variables. Conversely, if the measurements have a lower time resolution
 than the causal influence, a model can be used in which the causal influences are contemporaneous or instantaneous 
 \citep{white2010}. 
 For details on structure learning from undersampled time series, see \cite{danks2013, gong15, plis2015}.

In multivariate time series literature, Structural vector autoregressive (SVAR) models are powerful tools for learning the structure of time series. 
SVAR allows causal influences to occur contemporaneously and with time lags. \cite{Swanson1997, demiralp2003, moneta2006, runge2019b}
 exploit constraint-based methods, such as  PC (Peter-Clark) \citep{spirtes1991}
  algorithm for the SVAR estimation. Such methods rely on Gaussianity and/or faithfulness assumption 
  (see Section~\ref{s:s2} for definitions). \cite{hyvarinen2010, moneta2013, dallakyan2020} 
   propose methods for non-Gaussian data. \cite{entner2010, malinsky2018} exploit the FCI algorithm to allow for the unmeasured confounding effects. \cite{chu2008} introduced additive
    non-linear time series models (ANLTSM) with linear contemporaneous effects for performing relaxed conditional independence tests. \cite{peters2013b} generalize ANLTSM and allow 
    for the non-linear contemporaneous effects in their time series 
models with independent noise (TiMINo) approach.
Recently, \cite{pamfil2020} propose a fully continuous optimization approach for learning the structure of time series by 
exploiting a novel characterization of acyclicity constraint introduced in \cite{zheng2018}.

In this work, we squarely depart from the time domain and propose a novel approach to recover
 a topological ordering of time series in the frequency domain.  For an overview of spectral dependence modeling
 in multivariate time series and its advantages, see \cite{ombaopinto2021}. 
 We name our procedure
 \textbf{Fre}quency \textbf{Dom}ain structure learning (FreDom). In sharp contrast to existing literature, which utilizes SCMs in the time domain,
  we define a complex-valued SCM in the frequency domain and study its close relation with the Cholesky decomposition 
  of the inverse spectral density matrix. In particular, for each frequency of the Fourier transform of the time series, 
  we establish a class of cSCMs. 
   Assuming that the time series is generated from the transfer function model, where the transfer functions are
    the inverse of the difference between identity matrix and weighted adjacency matrix at each frequency 
    (see Section~\ref{s:sSEM} for details), we estimate the ordering of
 the summary DAG from the cSCM. Given the ordering, we propose a regularized likelihood approach to recover the summary DAG. In addition, we also provide an extension of our approach that relies on the complex-valued formulation of NOTEARS \citep{zheng2018}
 to estimate the summary DAG.
To the best of our knowledge, 
the only frequency domain approach for learning the structure of time series is proposed in   \cite{shajarisales15,besserve2021causeeffect},
but the latter is limited only to cases when the number of series is equal to two.   

Compared to the existing methods, another important advantage of FreDom is that it 
allows to work with a complex-valued time series or 
sequence data. The latter is naturally used in  telecommunications, robotics, bioinformatics, image processing, radar, and speech recognition \citep{schreierscharf2010,wolter2018complex,Wolter2018FourierRF,yangmali2020,chiyanetal2022}.

 Throughout the paper, we use the following notation: scalars are denoted by lowercase letters,
  except when they indicate the length of time series or frequency.
   To distinguish a (random) vector from a matrix, we highlight the former in bold. The dependence
    of vector or matrix from the time (frequency) index is represented by $\X(t)$ and $B(t)$, respectively, and 
the $ i$th element of the vector $\X(t)$ is denoted by $X_i(t)$. The conjugate, and the conjugate transpose of the complex-valued
matrix is denoted by $\bol B^{*}$ and $\bol B^{H}$, respectively. In addition, we place all appendices in Supplementary Materials.

\section{Methods}
\label{s:s2}

We start by reviewing the existing literature on structure learning for \textit{iid} data.
The goal of structure learning is to recover the underlying structure of variables $X_i,\, i \in E$,
  given the samples from the distribution $\Pro$. We let $\Gi(V,E)$ be a directed acyclic graph (DAG) on $E$ that describes the relationship between variables.
  Independence-based (also called constraint-based) methods \citep{spirtes1991, pearl2009}, score-based
methods \citep{heckerman1995, Chickering2002, teyssier2005, loh2014},
and functional-based methods \citep{shimizuetal06a, petersetal2014,zhangetal2015, chen2019} are three popular approaches to learning
the structure of the underlying DAG.

Independence-based methods, such as the
inductive causation (IC) \citep{pearl2009} and  PC (Peter-Clark) \citep{spirtes1991} algorithm, utilize conditional
independence tests to detect the existence of edges between each pair of variables. The method assumes that the
distribution is Markovian and faithful for the underlying DAG, where $\Pro$ is faithful to the DAG
$\Gi$ if all conditional independencies in $\Pro$ are entailed in $\Gi$, and Markovian if the factorization property 
$\Pro(X_1,\dots,X_p) = \prod_{j=1}^{p}\Pro(X_j| \Pi^{\Gi}_{j})$ is satisfied. Here  $\Pi^{\Gi}_{j}$ is the set of all
 parents of a node $j$. 
 In contrast to constraint-based methods, the score-based approach treats structure learning 
as a combinatorial optimization problem. In particular, in the DAG space, they search and 
test various graph structures by assigning a score to each graph and selecting
the one that best fits the data. Finally, the functional-based methods restrict the functional 
class and the error term distributions so as to achieve identification.  
 
 \subsection{Bayesian Networks and SCM} \label{s:SEM}
 The SCM for a random vector $\X = \{X_i|i \in E\}$ is a 4-tuple $(\X, \varep, \mathcal{F}, P(\varep))$, where
 $\varep$ is a set of background (exogenous) variables, $\mathcal{F}$ is a set of functions $\{f_1,f_2, \dots, f_p\}$  where each
 $f_i$ maps $\varepsilon_i \cup \Pi^{\Gi}_{i}$ to $X_i$, and $P(\varep)$ is a probability function defined over the domain of $\varep$. 
  SCM posits casual relations, such that for all $i \in E$, $X_i := f_i(\Pi^{\Gi}_{i}, \varepsilon_i)$, where $\varepsilon_i,i \in E$ are jointly
  independent and the causal structure is encoded in a DAG $\Gi$ \citep{pearl2009,bareinboim2020}. 

\noindent
   For example, if  $f_i$s
   are linear and have additive noise, SCMs can be written as 
 \begin{equation} \label{eq:sem}
 X_j := \sum_{k \in \Pi^{\mathcal{G}}_j} \beta_{jk} X_k + \varepsilon_j, \; j = 1, \dots, p,
 \end{equation}
 Denoting the weighted adjacency matrix $B = (\beta_{jk})$ with zeros along the diagonal,
 the vector representation of (\ref{eq:sem}) 
 \begin{equation} \label{eq:vsem}
 \X := B \X + \varep,
 \end{equation}
 where $\varep: = (\varepsilon_1, \dots, \varepsilon_p)^{'}$ and $\X: = (X_1, \dots, X_p)^{'}$.
A DAG admits a topological ordering $\varrho(\cdot)$ with which a $p \times p$ permutation matrix $P_{\varrho}$ can be associated such that
  $ P_{\varrho}\bol x = (x_{\varrho(1)}, \dots, x_{\varrho(p)})$, for $\bol x \in R^p$. The existence of a topological order leads to the permutation-similarity of $ B$ 
  to a strictly lower triangular matrix $ B_{\varrho} =  P_{\varrho} BP^{'}_{\varrho}$ by permuting rows and columns of $B$, respectively \citep{bollen1989}. 

\subsection{Complex-Valued Bayesian Networks and cSCM}
\label{s:cSEM}

We define $\bol Y \in C^p$, be iid complex-valued, proper random vectors. 
The complex-valued SCM and corresponding DAG $\Gi$ can be defined analogously to real-valued SCM by
\begin{equation} \label{eq:csem}
	\bol Y := f(\bol Y, \varep_c)
\end{equation} 
For example, for linear Gaussian BN $\bol Y \sim N_c(0, \Sigma_c)$, then 
$E[\bol Y \bol Y^H] = \Sigma_c \in C^{p \times p} = \sigma^2(I - B)^{-1}\{(I-B)^{H}\}^{-1}$, and 
the weighted adjacency matrix $B \in C^{p \times p}$ 
 is potentially complex-valued where the subscript $c$ indicates that the 
distribution is complex-valued and $A^H$ denotes the conjugate transpose $(A^{*})^{'}$. For details on complex-valued Gaussian distribution, see Chapter 2 in  \cite{andersen1995}.
 
\section{Complex-Valued Bayesian Networks For Time Series} \label{s:sSEM}

We now return to structure learning for time series, given $\Xt \in R^p$ or $C^p$ for $t = 1,\dots,T$ such that 
the autocovariance function satisfies $\sum_{h = -\infty}^{\infty}{|\gamma(h)| < \infty}$, i.e. the spectral density matrix exists \citep{Brockwell1986}.  
Recall that the discrete Fourier transform (DFT) for the time series $\Xt$ is 
\begin{equation} \label{eq:dft}
	\bol d(\omega_k) = \frac{1}{\sqrt{T}} \sum_{t = 1}^{T} \Xt \mbox{exp}(-2\pi i  \omega_k t),
\end{equation}
$\bol d^*(\omega_k) = \bol d(-\omega_k) = \bol d(1-\omega_k)$ and from \cite[Theorem 4.4.1]{Brillinger2001} as 
$T \rightarrow \infty$, $\bol d(\omega_k),\; k = 2,3, \dots, (T/2)-1$ are independent complex Gaussian 
$N_c(0, S(\omega_k))$ random vectors and for $k = \{1, T/2,T\}$, $\bol d(\omega_k)$ are independent real Gaussian
 $N_r(0, S(\omega_k))$, where $S(\omega_k)$ is the spectral density
matrix at the Fourier frequency $\omega_k$. We assume that the DFT $\bol d(\omega_k)$ satisfies the cSCM with the additive error
at each Fourier frequency $\omega_k,\; k = 1, \dots, T/2$:
\begin{equation} \label{eq:fcsem}
    \bol d(\omega_k)  = f(\bol d(\omega_k)) + \varep(k).
\end{equation}
We denote the adjacency matrix of the graph $G$ by $W$, where $W_{ij}= 1$ if 
$ d(\omega_k)_j \rightarrow d(\omega_k)_i$. Note that if $f$ is linear
then the coefficient matrix $B$ has the same non-zero pattern as $W$.

Next we define a \textbf{summary} DAG for the frequency domain.
\begin{definition} \label{d:sdag}
A summary DAG $\Gi$ for the time series $\Xt$ is a DAG which has an arrow from
 $\X(t)_i$ to $\X(t)_j$, $i \neq j$, if 
$W_{ji}(\omega_k) \neq 0$ for some $k = 1, \dots, T/2$.
\end{definition}
 For the next section we impose the following structure invariance assumption on DFT and time series.
\begin{assum} \label{as:as1} (Structure Invariance)
The structure of time series $\Xt$ and DFT $\bol d(\omega_k),\, (k,t = 1, \dots, T)$   remains unchanged across the time and frequency points.
\end{assum}

Discussion of the relaxation of assumption~\ref{as:as1} is provided in Section~\ref{s:recdag}.

\subsection{Linear Case} \label{s:caselin}
In this section, we assume $f$ is linear in (\ref{eq:fcsem})
\begin{equation} \label{eq:lcsem}
	\bol d(\omega_k)  := B(\omega_k) \bol d(\omega_k) + \varep(k).
\end{equation}
where $B(\omega_k) \in C^{p \times p}$ entails the underlying structure of the summary DAG. Consequently, from the inverse Fourier
 transform and (\ref{eq:lcsem}), \textbf{the time series is generated} from the transfer function model
\begin{equation} \label{eq:ts}
\Xt = \sum_{k = 1}^{T}(I_p - B(\omega_k) )^{-1} \exp(2\pi i \omega_kt)\varep(k),
\end{equation}
where $i = \sqrt{-1}$, $\omega_k = k / T, \; k = 1, \dots, T$ and $\varep(k)$ are independent 
$N_c(0, (1/T)I_p),\, \varep(k) = \varep^{*}(T-k)$ for $\omega_t \neq \{0,0.5,1\}$, and real 
Gaussian $N_r(0, (1/T)I_p)$ otherwise. Moreover, from (\ref{eq:fcsem}), the spectral density matrix can be estimated by

\begin{equation} \label{eq:sden}
S(\omega_k) = \frac{1}{T}(I_p - B(\omega_k) )^{-1}\{(I_p - B(\omega_k) )^{-1}\}^{H}.
\end{equation}
A point of departure for our algorithm is an important
result for the real-valued SCM, which state that the graph $\Gi$ and the parameters $B$ can be identified from the covariance matrix under equal variance and causal sufficiency assumptions
\citep{peters2013}.
 \cite{ ghoshal2018, chen2019}
  observe that the ordering of certain conditional variances implies the identifiability of parameters.  Consequently, by ordering the estimates of
  those variables, the authors establish a fast method to learn the topological ordering of the variables. 
Next Lemma, which is the extension of  \cite[Lemmas~1]{chen2019} to 
 cSCM defined in
 (\ref{eq:lcsem}), is used to recover such topological ordering for cSCM.
 The proof is provided in Appendix~\ref{a:proof}, located in Supplementary Materials, for completeness.
 
 \begin{lemma} \label{l:eqvar}
 Let $\bol Y \in C^{p}$ is generated as in (\ref{eq:lcsem}). If the parent set
 $\Pi^G_j = \emptyset$ then $\mbox{var}(\bol Y_j) = 1/T$, otherwise
 $\mbox{var}(\bol Y_j) \geq 1/T * (1 + \eta) > 1/T$, where $\eta = \min_{(k,j) \in E} \beta_{jk}\beta_{jk}^{*}$.
 \end{lemma}

  The findings in Lemma~\ref{l:eqvar} allow to modify
  \cite[Algorithm~1]{chen2019} to complex-valued case, where at each Fourier frequency, the topological ordering of the  Fourier transform $\bol d(\omega_k)$ is 
  estimated by iteratively selecting a source node by comparing variances conditional on the previously
  selected variables. The main difference between FreDom and \cite{chen2019} is that in each frequency point the conditional variances
  are obtained from the (inverse)spectral density matrix, instead of covariance matrix. The Algorithm~\ref{a:fredom} summarizes the main steps.

\begin{algorithm}[ht!]
\caption{Stage 1 of FreDom Algorithm}
\label{a:fredom}
\begin{algorithmic}
\STATE \bfseries{Input}:
\STATE $\textit{$M$} \gets \textit{number of Fourier frequency points}$
\STATE $\textit{$S(\omega_k),\,k =1,\dots,M$} \gets \textit{spectral density matrix}$
\STATE $\Theta \in R^{M\times p} \leftarrow \emptyset$
\FOR {$k = 1$ {\bfseries to} $M$} 
\FOR  {$i = 1$ {\bfseries to}  $p$} 
\STATE  {$\theta \leftarrow \argmin_{j \in V/\Theta[k,i-1]}f(S(\omega_k), \Theta[k,(i-1)], j)$}
\STATE   $\Theta{[k,i]} = \theta$
\ENDFOR
 \ENDFOR
\STATE {\bfseries Output}:$\mbox{ the most commonly occurring row of } \Theta.$
\end{algorithmic}
\end{algorithm}
Here, each row of matrix $\Theta$ stores estimated topological ordering in each Fourier frequency.
 It is instructive to note that from Assumption~\ref{as:as1}, 
 $B(\omega_k)$ has the same structure for $k =1, \dots, T/2 - 1$, that is, the same zero patterns.
 However, due to sampling variability in the observed time series, the estimated ordering of variables in (\ref{eq:fcsem})  may vary at some Fourier frequencies.
  Therefore, we choose the ``best'' estimated order of the summary DAG as the most common among the Fourier frequencies. 
  It is important to note that if there is a priori knowledge of the importance of a particular frequency interval, for example, 
lower frequencies, then different importance weights can be applied to the frequencies to select the order of variables.

To select a source node by comparing conditional variances, in Algorithm~\ref{a:fredom}, we minimize the frequency domain 
analog of \cite{chen2019} criterion
\begin{equation} \label{eq:ffunc}
f(S(\omega_k), \Theta[k,(i-1)], j) =  (\hat S(\omega_k))_{j,j} -  (\hat S(\omega_k))_{j, \Theta} (\hat S(\omega_k))^{-1}_{\Theta,\Theta}  (\hat S(\omega_k))_{\Theta,j}.
\end{equation}
 

 \subsection{Recovering DAG from topological ordering} \label{s:recdag}

In the first stage of FreDom, Algorithm~\ref{a:fredom} returns the topological ordering of a summary DAG.
In the Stage 2 of FreDom, we recover the summary DAG from the frequency domain topological ordering. 
As discussed in Section~\ref{s:SEM}, given a topological ordering $\varrho$, $B_{\varrho}$ is lower triangular. Similarly,
from (\ref{eq:fcsem}) and (\ref{eq:sden}), given the ordering, $B_{\varrho}(\omega_k)$ is lower triangular, and 
$L_{\varrho}(\omega_k) = \sqrt{T}(I - B_{\varrho}(\omega_k))$ is the Cholesky factor of the inverse spectral density
 matrix $\Omega_{\varrho}(\omega_k) = S_{\varrho}^{-1}(\omega_k) = L_{\varrho}^H(\omega_k)L_{\varrho}(\omega_k)$. 
 From now on, whenever there is no confusion, we drop the subscript $\varrho$.
 Ignoring $k = \{1, T/2\}$ frequency points, from (\ref{eq:dft}),  the joint pdf for $\bol d(\omega_k),\, k = 2, \dots, (T/2) - 1$ is

\begin{equation}\label{eq:pdf}
\begin{aligned}
g(\bol d(\omega_2),\dots, \bol d(\omega_{(T/2) - 1})) = 
\prod_{k = 2}^{(T/2) - 1} \frac{\mbox{exp}(- \bol d^H(\omega_k) L^H(\omega_k)L(\omega_k) \bol d(\omega_k))}{\pi^p \mbox{det}(S(\omega_k))}
\end{aligned}
\end{equation}

\noindent
A standard assumption in spectral density estimation is locally smoothness \citep{Brillinger2001, stoica1997}, i.e., $S(\omega_k)$
 is approximately constant over $N = 2m_t + 1$ consecutive frequency points where $m_t$ is the half-window size. After carefully picking

\[\begin{aligned} \tilde \omega_k &= \frac{(k - 1)N + m_t + 1}{T};\;  \\
M &=  \Big \lfloor \frac{T/2 - m_t - 1}{N} \Big \rfloor;\; k = 1,2, \dots M,
\end{aligned}\]

\noindent
leads to $M$ equally spaced frequencies $\tilde \omega_l$. Therefore, the exploitation of the local smoothness assumption results for $l = -m_t, -m_t + 1, \dots, m_t$
\begin{equation} \label{eq:smooth}
\tilde \omega_{l,k} = \frac{(k - 1)N + m_t + 1 + k}{T};\, S(\omega_k) = S(\omega_{\{l,k\}}).
\end{equation}
From  (\ref{eq:pdf}) and (\ref{eq:smooth}), the pdf is
\begin{equation} \label{eq:modpdf}
\begin{aligned}
g(\bol d(\omega_2),\dots, \bol d(\omega_{(T/2) - 1})) 
&= \prod_{k = 1}^{M} \prod_{l = -m_t}^{m_t}  \frac{\mbox{exp}(-\bol d^H(\tilde \omega_{l,k}) L^H(\omega_k)L(\omega_k) \bol d(\omega_{l,k}))}
{\pi^p \mbox{det}(S(\omega_k))^N} \\
&= \prod_{k = 1}^{M} \frac{\mbox{exp} \{-N\mbox{tr}(\tilde S(\omega_k)L^H(\omega_k)L(\omega_k) )\}}{\pi^{Np}  \mbox{det} (L^H(\omega_k)L(\omega_k))^{-N}},
\end{aligned}
\end{equation}
where $\tilde S(\omega_k) = \sum_{l= -m_t}^{m_t} \bol d(\tilde \omega_{l,k}) \bol d^H(\tilde \omega_{l,k}) / N$
 is the sample spectral density matrix whose entries are potentially complex-valued. Thus, the log-likelihood function can be written as

\[ \begin{aligned}
W(L[\cdot]) = \sum_{k = 1}^{M} N [\log \mbox{det} (L^H(\omega_k)L(\omega_k)) 
 - \mbox{tr}(\tilde S(\omega_k) L^H(\omega_k)L(\omega_k)) ].
\end{aligned}\]
From Assumption~\ref{as:as1}, $B(\omega_k)$, and therefore $L(\omega_k)$, have the same structure over 
$k = 1, \dots, M$ frequency point. Thus, to impose a structure similarity assumption on the Fourier frequency points, we define the following constrained optimization problem 

\begin{equation} \label{eq:tslog}
\begin{aligned}
\min_{L[\cdot],Z} \quad & -W(L[\cdot]) +   P(Z, \lambda), \\
\mbox{s.t.} \quad & L(\omega_k) = Z, \, k = 1,\dots, M, 
\end{aligned}
\end{equation}
where
\begin{equation} \label{tsggl:1}
\begin{aligned}
P(Z, \lambda) &=  \lambda \sum_{ij} |Z_{ij}| \\  L[\cdot] &= \{L(\omega_1), \dots, L(\omega_M)\}.
\end{aligned}
\end{equation}

The constraints $L(\omega_k) = Z,\, k = 1, \dots, M$ is used to ensure that Assumption~\ref{as:as1}
is satisfied, i.e., in each Fourier frequency 
the summary DAG structures are the same and the penalty $P(Z, \lambda)$ introduces sparsity.  
The minimization problem (\ref{eq:tslog}) is convex, and the existence of a minimizer is guaranteed for any choice of 
$\lambda \geq 0$ \citep[Theorem 27.2]{rockafellar1970}. 
We appeal to the ADMM (alternating direction method of multipliers) algorithm for minimizing
 (\ref{eq:tslog}) \citep{Boyd2011, dallakyan2021c, ngzhang2022}. The ADMM minimizes the scaled
 augmented Lagrangian 
\begin{equation} \label{eq:14_0}
\begin{split}
\mathcal{L}_\rho(\Theta[\cdot],Z,U[\cdot])&= \sum_{n=1}^{M} N [ -\log \mbox{det} (L^H(n)L(n)) + \mbox{tr}(\tilde S(n) L^H(n)L(n))] \\
& +{\rho}\sum_{n =1}^{M} (\|L(n)-Z+U(n)\|_F^2 - \|U(n)\|_F^2)+ P(Z, \lambda),
\end{split}
\end{equation}
where $\rho>0$ is the penalty coefficient, $U(n),\, n = 1,\dots,n$ are the Lagrangian 
multipliers, and $\|X(n)\|^2_F = \sum_{i j} |X_{ij}(n)|^2$.
Given $(L^{(k)}[\cdot],Z^{(k)},U^{(k)}[\cdot])$  matrices in the $k$th iteration, the ADMM algorithm implements the following three updates for the next ($k +1 $) iteration:
\begin{enumerate}
	\item[(a)]\label{step1} 
      $	L^{(k+1)}[\cdot] \leftarrow \argmin_{L[\cdot]} \mathcal{L}_\rho(\Theta[\cdot],Z^{(k)},U^{(k)}[\cdot])$
	\item[(b)] \label{step2}
	$Z^{(k+1)} \leftarrow  \argmin_{Z]}  \mathcal{L}_\rho(L^{(k+1)}[\cdot],Z,U^{(k)}[\cdot])$
	\item[(c)] \label{step3} $U^{(k+1)}[\cdot] \leftarrow U^{(k)}[\cdot]+(L^{(k+1)}[\cdot]- Z^{(k + 1)})$
\end{enumerate}

\noindent
Interestingly, as we show in Appendix~\ref{a:steps13}, each of the updates (a)-(b) have closed form solutions. 
 Moreover, in contrast to real-valued ADMM formulation, where $L$ is real-valued, in (\ref{eq:tslog}) $L[\cdot]$ is complex-valued.  To solve complex-valued optimization, we resort to  Wirtinger calculus \citep{wirtinger1927,brandwood1983},
   together  with the definition of Wirtinger subgradients \citep{bouboulis2012}. 

\noindent
It is timely to note that the choice of Fourier frequency points $M$ depends on the half-window 
size $m_t$ and there exist data-driven approaches such
as gamma-deviance-GCV \citep{ombaoetal2001} for automatic selection.

\paragraph*{Violation of Assumption~\ref{as:as1}.} \label{s:discuss1}

In some datasets, it is expected to observe violation of Assumption~\ref{as:as1}.
For example, data obtained from the brain
signals may follow locally stationary process \citep{dahlhaus1997} or exhibit change point behavior over the 
frequency points (for details, see \cite[Chapter 7]{ombaopinto2021}). In such cases, it is expected to see a change 
in the domain structure. Our proposed framework gracefully handles such structure changes by adopting fused lasso or $\ell_1$-trend penalties \citep{dallakyan2022}, instead of  
(\ref{tsggl:1}). In particular, by enforcing for $1 \leq i,j \leq p$  

$$P(L[\cdot], \lambda) = \lambda \sum_{n = 2}^{M} [L_{ij}(n) - L_{ij}(n-1)],  $$
we expect to detect the structural changes.

\paragraph*{Including lag effect.} \label{s:discuss2}
Another limitation of formulation (\ref{eq:lcsem}) is that it does not model 
the effect of lags in the frequency domain (for example, see \cite[Example 3]{ombaopinto2021}).
For the real-valued problem, \cite{pamfil2020} rely on 
a novel acyclicity constraint introduced in \cite{zheng2018} (NOTEARS) to estimate
a SVAR model. As we show in Section~\ref{s:fredomext}, with some careful modification,
NOTEARS can be applied to the FreDom framework. Consequently, the lag effect 
can be modeled in (\ref{eq:lcsem}) by extending \cite{pamfil2020} framework.


\paragraph*{Summary DAG vs Summary Graph.}

\begin{figure}[ht!]
\centering
 \begin{tikzpicture}[
            > = stealth, 
            shorten > = 1pt, 
            auto,
            node distance = 2cm, 
            semithick 
        ]

        \tikzstyle{every state}=[
            draw = black,
            thick,
            fill = white,
            minimum size = 1.3cm
        ]


        \node[state] (Y1) {$y_{1t-1}$};
        \node[state] (Y2) [above of=Y1] {$y_{2t-1}$};
        \node[state] (Y3) [below of=Y1] {$y_{3t-1}$};
       \node[state] (Y11) [right of=Y1] {$y_{1t}$};
        \node[state] (Y22) [above of=Y11] {$y_{2t}$};
        \node[state] (Y33) [below of=Y11] {$y_{3t}$};

  
        \node[state] (Y1t) [right of =Y11] {$y_{1t}$};
       \node[state] (Y3t) [below of=Y1t] {$y_{3t}$};
        \node[state] (Y2t) [above of=Y1t] {$y_{2t}$};

	
        \node[state] (Y1d) [right of =Y1t] {$y_{1t}$};
       \node[state] (Y3d) [below of=Y1d] {$y_{3t}$};
        \node[state] (Y2d) [above of=Y1d] {$y_{2t}$};


        \path[->] (Y2) edge node {}(Y1);
        \path[->] (Y1) edge node {}(Y3);
        \path[->] (Y2) edge node {}(Y22);
        \path[->] (Y2) edge node {}(Y11);
        \path[->] (Y1) edge node {}(Y33);
        \path[->] (Y3) edge node {}(Y11);
        \path[->] (Y22) edge node {}(Y11);
        \path[->] (Y11) edge node {}(Y33);

        
         \path[->] (Y1t) edge node {}(Y3t);
        \path[->] (Y2t) edge node {}(Y1t);
        \path[->] (Y3t) edge node {}(Y1t);

        
        \path[->] (Y2d) edge node {}(Y1d);
        \path[->] (Y1d) edge node {}(Y3d);
    \end{tikzpicture}
	\caption{Illustration of (Left) Full, (Middle) Summary Graph And (Right) Summary DAG 
	For The Linear SVAR Where The Number Of Lags $q = 1$ and $p = 3$.}
	\label{fig:exdag}
\end{figure}
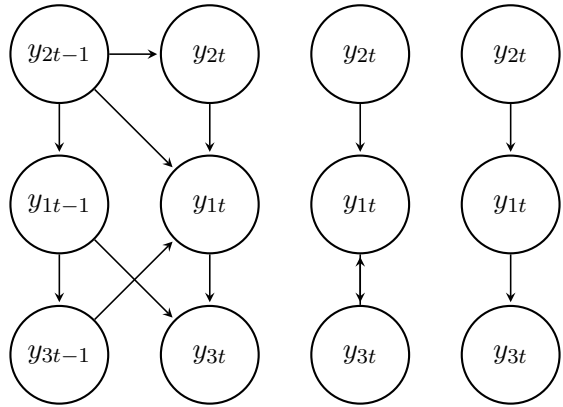

\noindent
FreDom algorithm always returns a DAG, which similar to the summary graph defined in \cite{peters2013b} does not provide explicit information about the
 structural relationship between lagged variables. Note that our definition of summary DAG is different
  from the summary graph defined in \cite{peters2013b}. Typically, the summary graph is not necessarily a  DAG and there is an arrow between $X_i(t)$ to $X_j(t)$, $i \neq j$,
  if there is an arrow from $X_i(t-k)$ to $X_j(t)$ in the full graph for some $k$. Figure~\ref{fig:exdag}(Middle) illustrates the example where the summary 
  graph is not a DAG and Figure~\ref{fig:exdag}(Right) shows the corresponding estimated summary DAG from the FreDom algorithm.

Interestingly, a summary DAG can be considered as a ``\textit{DAG projection}'' of the summary graph. To motivate this
 claim, we assume a time series that is generated from the linear SVAR with contemporaneous effects.
\begin{equation} \label{eq:linsvar}
\begin{aligned}
	\Xt = &B_0\Xt  + B_1 \Xto + \bol\varepsilon(t)  \\= &(B_0 + B_1R)\Xt +  \bol\varepsilon(t),
\end{aligned}
\end{equation}
where $R$ is a backshift operator $\Xto = R\Xt$, $B_0$ and $B_1$ encompass the contemporaneous and lagged structure. 
From (\ref{eq:linsvar}) and (\ref{eq:dft}), we can write \citep{wei2006}
\begin{equation}
\bol d(\omega_k) = [B_0 + B_1\mbox{exp}(-i\omega_k)] \bol d(\omega_k) + \bol \varepsilon(k).
\end{equation}
 Comparing the latter with the (\ref{eq:fcsem}), the claimed relationship
of the summary graph and summary DAG follows.
%

\subsection{FreDom extension} \label{s:fredomext}
In Section~\ref{s:caselin} we imposed two limitations on (\ref{eq:fcsem}): linearity and
equal variance. In this section, we show flexibility of our approach by extending
FreDom to more general cases using NOTEARS framework,
proposed in \cite{zheng2018, zheng2020}. We name the \textbf{Ex}tended
FreDom framework as ExFreDom. 

\noindent
Recall that NOTEARS formulates structure learning of SCM as a continuous constrained optimization
problem

\begin{equation} \label{eq:notear}
\begin{aligned}
\min_{B} \quad & \ell(B;\bol Y) +   P(B, \lambda), \\
\mbox{s.t.} \quad & h(B) = \mbox{tr}(e^{B \odot B}) -p = 0, 
\end{aligned}
\end{equation}
where $\ell(B;\bol Y)$ is the least-squares loss, $P(B, \lambda)$ is defined in 
(\ref{tsggl:1}), and $h(B) \geq 0$ is the acyclicity constraint, which is equal
0 if and only if $B$ is a DAG. 

\noindent
Note that (\ref{eq:notear}) is not directly extendable to complex-valued case, since
when $\bol Y \in C^{p \times p}$ then $B$ is complex-values and $B \odot B$ is not 
non-negative matrix. For cSCM, we define $h(B) = \mbox{tr}(e^{B \odot B^{*}}) -p$,
where $B^{*}$ is a complex conjugate and $h(B) \geq 0$. Thus, to learn the summary DAG of
$\Xt$ in the frequency domain, we solve the following optimization problem, formulated
as an ADMM problem

\begin{equation} \label{eq:fredomext}
\begin{aligned}
\min_{B[\cdot],Z} \quad & \sum_{n = 1}^{M}\ell(B(n);\bol d(n)) +   P(Z, \lambda), \\
\mbox{s.t.} \quad & h(Z) = 0, \\
\quad &B(k) = Z,\, k = 1, \dots, M.
\end{aligned}
\end{equation}

\noindent
As in case of (\ref{eq:tslog}), the constraints $B(k) = Z,\, k = 1, \dots, M$ is used to ensure that Assumption~\ref{as:as1} is satisfied, i.e., in each Fourier frequency 
the corresponding structures are the same. We note that a similar, real-valued formulation is given
in \cite{ngzhang2022} to solve federated learning problem. Details of the solution
of (\ref{eq:fredomext}) and its implementation is given in Appendix~\ref{a:steps19}.

\section{Numerical Experiments} \label{s:num}
In this section, we illustrate the performance of FreDom and ExFreDom
on complex and real-valued time series data.
In addition, we consider two real-world applications: Air Pollution and Stock Return Volatility Data which corroborate the simulation results.
For the real-valued simulation analysis we compare the performance of our methodology with DYNOTEARS \citep{pamfil2020} and VARLINGAM \citep{hyvarinen2010}.
Additional simulation analyses can be found in Appendix~\ref{a:addexp}.

\subsection{Simulated Data}
For all experiments, the length of the time series is $T = 1000$, and all simulations are repeated 50 times. 
One challenge in adopting the frequency domain approach is that it requires parameter estimation at each 
Fourier frequency. For example, for $p$ dimensional time series and $M$ Fourier frequencies,
FreDom estimates $Mp (p+1)/2$ parameters.
Based on simulations, choosing  $M = (5,10)$ gives satisfactory results.  

Performance is measured using the structural hamming distance (SHD) and Structural Intervention Distance (SID) \citep{petersbuhlmann2015}. 
The SID quantifies the proximity between two DAGs in terms of their respective causal inference statements.
 A lower value of SID and SHD indicates a better performance.
 For real data analysis, we also compare FreDom with Granger causality.


 The tuning parameter for FreDom is
selected using extended BIC \citep{foygel2010}.  In particular, we define a grid of search space
$[\lambda_{min}, \lambda_{max}]$, where
$\lambda_{min}$ and $\lambda_{max}$ chosen such that to avoid very dense and sparse models, respectively.
We start by finding $\lambda^{*}$ that results to the graph with no edges and choose $\lambda_{max}= \lambda^{*}/ 2$, to avoid
 very sparse models. In addition, we use ``warm''-starting strategy over the grid for a faster convergence.
 
\begin{figure}
	\vskip 0.2in
	\centering
	\includegraphics[width = 0.65\columnwidth, height = 8.5cm]{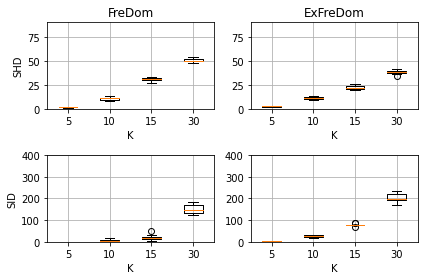}
	\caption{SHD and SID metrics for Experiment 1.}
	 \label{f:exp1}
 	\vskip -0.2in
\end{figure}

 
\paragraph{Experiment 1: Complex-valued time series.} We utilize
\cite[Theorem 1]{dai2004} (for details, see Appendix~\ref{a:exp1}),
which states that (\ref{eq:ts}) can be used to generate a complex-valued
time series whose 
topological order and spectrum are identical to the given order and spectrum at Fourier frequencies, to simulate complex-valued time series from the given random summary DAG 
for $K = \{5,10,15,30\}$.
 For each Fourier frequency $\omega_k$, we construct the Cholesky factor of the inverse spectral density by performing the following steps: (1) 
 Fix the order and fill the adjacency matrix with zeros, (2)  Replace every matrix entry in the lower triangle (below the diagonal) by 
 independent realizations of Bernoulli(s) random variables with success probability $\mathit{s}$, $0 <\mathit{s} < 1$, where $\mathit{s}$ 
 reflects the sparseness of the model. We select $s = 0.2$ for this experiment. (3) Finally,  in the adjacency matrix replace each entry with a $1$
  by the independent realizations of a  $c_1 \cos(4\pi  \omega_k) + 1.2i  c_2\sin(2\pi \omega_k)$, where $c_1, c_2$ are randomly selected from
   the $U[-0.1,-1] \cup [0.1,1]$ distribution. The above procedure ensures that the DAG structure 
of the generated time series is the same for all frequencies, and $B(\omega_k)$ in (\ref{eq:fcsem}) is only a function of $\omega_k$. 
Figure~\ref{f:exp1} presents the results. We can see that when $K =5$, the SID and SHD metrics are close to 0, with very small variance
and, as expected, the performance deteriorates as $K$ increases.  

\begin{figure}[ht!]
 	\vskip 0.2in
	\centering
	\includegraphics[width = 0.6\columnwidth, height = 4.5cm]{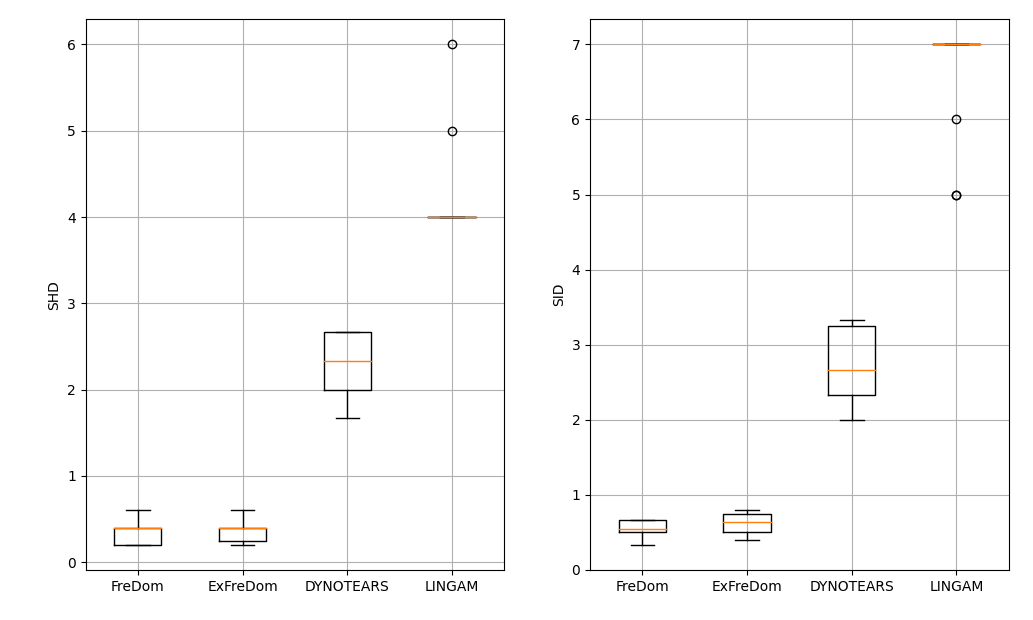}
	\caption{SHD and SID metrics for Experiment 2.}
	 \label{f:exp4}
	 \vskip -0.2in
\end{figure}

\paragraph{Experiment 2: Data from the non-linear SVAR Model.} 
Similar to \cite{peters2013b}, we simulate dataset from 
$X_1(t) = b_{11} X_2(t)^2 + b_{12} X_1(t-1) + b_{13}X_2(t-1)^2 +  u_1(t),\, 
X_2(t) = b_{22}X_2(t-1)  + u_2(t), \, 
X_3(t) = b_{31}X_1(t)^3 + b_{32}X_2(t-1)^2 + b_{33}X_3(t-1) + u_3(t),\, 
X_4(t) = \mbox{exp}(b_{41}X_{3t}) + b_{42}X_4(t-1) + u_4(t)$
, where
$u_i(t) \sim N(0,1)$ and $b_{ij}\sim U[-0.1, -0.4] \cup [0.1,0.4]$. Figure~\ref{f:exp4} 
shows the simulation results. 
As we can see, FreDom and ExFreDom report the best results.

\subsection{Air Pollution Data}
\label{s:air}

We use (Ex)FreDom to estimate a summary DAG for 5 time series of air pollutants of length 8370. The series were recorded hourly during the year 2006 at Azusa, California. Data can be obtained from the Air Quality and Meteorological Information System. Recorded variables 
include CO and NO (pollutants mainly emitted from the cars and industry), $\mbox{NO}_2$ and $\mbox{O}_3$ (generated from different reactions in
the atmosphere), and the global solar radiation intensity $\mbox{R}$. The similar datasets were analyzed in \cite{dahlhaus2003} and \cite{Davis2016}.

Figure~\ref{f:aveplot} in Appendix~\ref{a:airadd} shows an average daily plot of five variables. Due to early morning traffic, 
CO and NO increase early, resulting in $\mbox{NO}_2$ increase. Higher $\mbox{NO}_2$ levels increase the Ozone ($\mbox{O}_3$) and the global radiation levels  throughout the day. 
 
\begin{figure}[!ht]
\centering
   \begin{subfigure}[b]{0.32\textwidth}
      \includegraphics[width=\textwidth,height = 3cm]{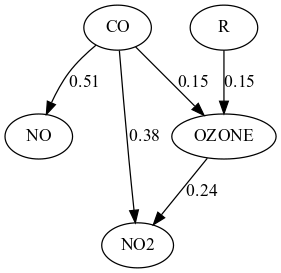}
      \caption{FreDom}
    \end{subfigure}
     \begin{subfigure}[b]{0.32\textwidth}
      \includegraphics[width=\textwidth,height = 3cm]{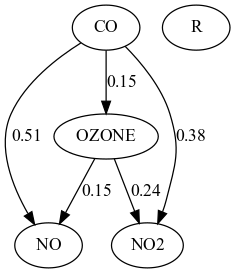}
      \caption{ ExFreDom}
    \end{subfigure}%
    \caption{The estimated DAG from the air pollution data.} 
\label{f:sumdag}
  \end{figure}

 Following \cite{dahlhaus2003}, we apply FreDom to the residual series after subtracting the  daily averages, as shown in 
Figure~\ref{f:aveplot}. The missing values in the original series are filled in by interpolating the residual series using splines. 
Figure~\ref{f:sumdag}(a) and Figure~\ref{f:sumdag}(b) report the estimated summary DAGs from FreDom and ExFreDom, respectively. 
 The weights on the edges report the absolute values of the partial spectral coherence, which are frequency domain analogues of partial correlations. 
Additional results for LINGAM, NOTEARS,
and Granger causality are available in Appendix~\ref{a:airadd}.

The summary DAG of FreDom correctly reflects the effect of global radiation R plays on $\mbox{O}_3$ generation. FreDom is also capturing the generation of $\mbox{NO}_2$
from $CO$ and the contemporaneous relation of $\mbox{CO}$ and $\mbox{NO}$ as the latter two pollutants are emitted from cars. However, we cannot 
validate the direction of the edge from the $\mbox{CO}$ to $\mbox{NO}$. The direction of the arrow from $O_3$ to $\mbox{NO}_2$ is reversed since 
$\mbox{O}_3$ is created from $\mbox{NO}_2$. Compared to FreDom, ExFreDom misses 
edge from R to $\mbox{O}_3$ and has an additional edge that correctly captures the
effect of $\mbox{CO}$ on $\mbox{NO}_2$.

\subsection{Stock Return Volatility Data}
\label{s:stock}

In this section, we analyze stock return volatility data. Data is taken from \cite{demirer2018}, where authors estimate the global bank network connectedness.
Original data contains 96 banks from 29 developed and emerging economies (countries) from September 12, 2003, to February 7, 2014.
 For illustration purposes, we select only economies where the number of banks in each economy is greater than 4, total of 54 banks (for more details, please refer to \cite{demirer2018}).

\begin{figure}[!ht]
	\vskip 0.2in
	\centering
	\includegraphics[width = 0.5\columnwidth, height = 5cm]{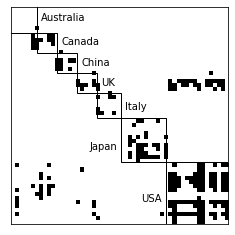}
	\caption{Estimated adjacency matrix with rows and columns sorted by country.}
	 \label{f:stockdag}
\end{figure}

Figure~\ref{f:stockdag} illustrates the estimated adjacency matrix from the FreDom algorithm. The rows and columns are sorted by country. 
As can be seen, banks from the same country 
tend to compose groups, meanwhile being connected to banks from the other countries. The latter result have been confirmed in many
macro-economic studies \citep{demirer2018}. The other interesting finding that needs more investigation is the causal relationship between UK and US banks. 
A similar result for the ExFreDom can be found in Appendix~\ref{a:stock}.



\section{Conclusion}
In this paper, we propose a frequency domain approach to recover the topological ordering of time series. 
Given the ordering, we propose a penalized likelihood approach to learn the summary DAG. 
The proposed algorithm effectively works for both real and complex-valued data.

For future work, we left the consideration of situations in which time series suffer from unobserved confounders or undersampling, 
as well as theoretical results on the probability bounds for recovering a true topological ordering of the summary DAG.

\bibliographystyle{tfnlm}
\bibliography{fredom}

\appendix

\section{Steps For (\ref{eq:tslog}) Minimization} \label{a:steps13}

We appeal to the alternating direction method of multipliers (ADMM)  \citep{Boyd2011} for minimization. The ADMM minimizes the scaled
 augmented Lagrangian 
\begin{equation} \label{eq:14}
\begin{split}
\mathcal{L}_\rho(\Theta[\cdot],Z,U[\cdot])&= \sum_{n=1}^{M} N [ -\log \mbox{det} (L^H(n)L(n)) + \mbox{tr}(\tilde S(n) L^H(n)L(n))] \\
& +{\rho}\sum_{n =1}^{M} \|L(n)-Z+U(n)\|_F^2 + P(Z, \lambda),
\end{split}
\end{equation}

Given $(L^{(k)}[\cdot],Z^{(k)},U^{(k)}[\cdot])$  matrices in the $k$th iteration, the ADMM algorithm implements the following three updates for the next ($k +1 $) iteration:
\begin{enumerate}
	\item[(a)]\label{step1_a} 
      $	L^{(k+1)}[\cdot] \leftarrow \argmin_{L[\cdot]} \mathcal{L}_\rho(\Theta[\cdot],Z^{(k)},U^{(k)}[\cdot])$
	\item[(b)] \label{step2_a}
	$Z^{(k+1)} \leftarrow  \argmin_{Z}  \mathcal{L}_\rho(L^{(k+1)}[\cdot],Z[\cdot],U^{(k)}[\cdot])$
	\item[(c)] \label{step3_a} $U^{(k+1)}[\cdot] \leftarrow U^{(k)}[\cdot]+(L^{(k+1)}[\cdot]- Z^{(k + 1)}[\cdot])$
\end{enumerate}

We use Wirtinger calculus to solve (\ref{eq:14}). For details on Wirtinger calculus, see \cite{Remmert1991, kreutz2009,brandwood1983}. 
It is instructive to note that for the update $L(\cdot)$, (\ref{eq:14}) can be separated into $M$ parallel problems. Thus, for $n = 1, \dots, M$ we solve

\begin{equation} \label{eq:14a} 
\begin{aligned}
N&[-\log \mbox{det}(L^{H}(n)L(n) + \mbox{tr}(\tilde S(n) L^{H}(n)L(n))] \\
 + &\rho \sum_{ij}(L_{ij}(n) - Z_{ij} - U_{ij}(n))^{*}(L_{ij}(n) - Z_{ij} - U_{ij}(n))
\end{aligned}
\end{equation}

Let $\bol \beta^i = (L_{ij}(n))^i_{j=1}$ denote the vector of lower triangular and diagonal entries in the $i$th row of $L(n)$, and $\tilde S_i$
denote the $i \times i$ submatrix of $\tilde S(n)$. Then after some algebra the following identities follow:

\begin{enumerate}
\item $\log \mbox{det}(L^{H}(n)L(n) = 2\sum_{i=1}^p \log \beta_i^i$
\item $\mbox{tr}(\tilde S(n)L^{H}(n)L(n)) = \sum_{i = 1}^p (\bol \beta^i)^{H}\tilde S_i \bol \beta^i$
\end{enumerate} 

Using the above results, (\ref{eq:14a}) can be written as a $p$ separate problems:

\begin{equation} \label{eq:14b}
\begin{aligned}
Q(\bol \beta^i) = \sum_{i = 1}^{p} & \Big \{ N[ (\bol \beta^i)^{H} \tilde S_i \bol \beta^i - 2 \log \beta^i_i] + \\
\rho & \sum_{j = 1}^{i - 1} (\beta^i_j - Z^i_{j} - U^i_{j})^{*}(\beta^i_j - Z^i_{j} - U^i_{j}) \Big\},
\end{aligned}
\end{equation}
where $Z^i$ and $U^i$ are the vector of lower triangular and diagonal values
in the $i$th row of $Z$ and $U$, respectively. We represent (\ref{eq:14b}) in a generic functional form
$h: C^{k-1} \times R_{+} \rightarrow R$

\begin{equation} \label{eq:14c}
h_{k, A, \rho} = N[-2 \log x_k + \bol x^h A \bol x] + \rho \sum_{j = 1}^{k -1}(x_j - z_j - u_j)^{*}(x_j - z_j - u_j),
\end{equation}
where $Q(\bol \beta^i) = h_{i, \tilde S_i, \rho}(\bol \beta^i)$ for $1 \leq i \leq p$.

\noindent
Since $h_{k,A,\rho}$ is convex, a sufficient and necessary condition for a global optimum
is 
$$\frac{\partial h_{k,A,\rho}}{ \partial x^{*}} = 0.$$

Using coordinatewise minimization algorithm and Wirtinger calculus, the following lemma follows.

\begin{lemma} \label{l:sol14c}
A minimizer of (\ref{eq:14c}) can be computed in a closed form.
\begin{equation}
x_j = \frac{\rho(z_j + u_j) -N( \sum_{l \neq j}A_{lj}x_l)}{N A_{jj} + \rho}
\end{equation}
for $1 \leq j \leq k -1$, and
\begin{equation}
x_k = \frac{- \mbox{Re}(\sum_{l \neq k}A_lk x_l) + \sqrt{\mbox{Re}(\sum_{l \neq k}A_lk x_l)^2 + 4 A_{kk}}}{2  A_{kk}}
\end{equation}
\end{lemma}
In Lemma~\ref{l:sol14c}, for complex number $a$, $\mbox{Re}(a)$ denotes the real part.

\paragraph{Update (b)}
From (\ref{eq:14}), the update of $Z$ is equivalent to solving complex Lasso problem.
In particular, a necessary and sufficient condition for a global minimum of $Z$ is that 
the subdifferential of $Q(Z) = {\rho}\sum_{n =1}^{M} \|L(n)-Z+U(n)\|_F^2 + P(Z, \lambda)$
must contain $\bol 0$

\begin{equation}
    \bol 0 \in - \sum_{n = 1}^{M}[L(n) - Z + U(n)] + \frac{\lambda}{\rho}\bol \Gamma,
\end{equation}
where the $ij$th component $\Gamma_{ij} = Z_{ij}/ |Z_{ij}|$ if $Z_{ij} \neq 0$
and $\Gamma_{ij} \in \{u: |u| \leq 1, u \in C\}$ otherwise. Thus, $ij$th component of update (b) solution is 

$$Z_{ij} = \frac{S_{\lambda/ \rho}(A_{ij})}{M},$$
where $A_{ij} = \sum_{n = 1}^{M} (L(n) + U(n))_{ij}$, and $S_{\lambda}(\cdot)$ is the soft-thresholding operator.

\section{Steps for (\ref{eq:fredomext}) Minimization} \label{a:steps19}
\noindent
Similar to NOTEARS, we use augmented Lagrangian
method to solve the optimization problem (\ref{eq:fredomext}), where the constrained 
optimization problem is reformulated as a sequence of unconstrained problems. For details,
see \cite{bertsekas2016}. For (\ref{eq:fredomext}), the augmented Lagrangian is given by

\begin{equation}
\begin{aligned}
    \mathcal{L}_{\alpha, \rho_1, \rho_2}(B[\cdot],Z, U[\cdot]) &= \sum_{n = 1}^M 
    \ell(B(n);\bol d(n)) +   P(Z, \lambda)+ \alpha h(Z) + \rho_1 h(Z)^2
    +\\ &\rho_2 \sum_{n =1}^{M} (\|L(n)-Z+U(n)\|_F^2 - \|U(n)\|_F^2),
\end{aligned}
\end{equation}

\noindent
Similar to (\ref{eq:tslog}), the ADMM algorithm implements the following four updates for the next ($k +1 $) iteration:
\begin{enumerate} 
	\item[(a)]\label{step1_1} 
      $	B^{(k+1)}[\cdot] \leftarrow \argmin_{B[\cdot]} \mathcal{L}_\rho(\Theta[\cdot],Z^{(k)},U^{(k)}[\cdot])$
	\item[(b)] \label{step2_1}
	$Z^{(k+1)} \leftarrow  \argmin_{Z}  \mathcal{L}_\rho(B^{(k+1)}[\cdot],Z,U^{(k)}[\cdot])$
	\item[(c)] \label{step3_1} $U^{(k+1)}[\cdot] \leftarrow U^{(k)}[\cdot]+(B^{(k+1)}[\cdot]- Z^{(k + 1)})$
	\item[(d)] \label{step4_!} $\alpha^{(k+1)} \leftarrow \alpha^{(k)} + \rho_1^{(k)}h(Z^{(k + 1)})$
	
\end{enumerate}

The updates of (a)-(d) are similar to \cite{ngzhang2022}, thus we omit the derivation
and only discuss the implementation details. Both NOTEARS-ADMM and NOTEARS-MLP-ADMM
algorithm in \cite{ngzhang2022} is implemented with PyTorch \citep{pazkeetal2019}
and use L-BFGS method to solve the unconstrained optimization problem in the augmented
Lagrangian method. Since PyTorch graciously handles Wirtinger calculus for complex-valued
data, at that end, no modification needed. However, L-BFGS is designed for real-valued data.
To adapt L-BFGS for the complex-valued data, we apply a well known trick \citep{Remmert1991}
where before passing the matrix $Z \in C^{p \times p}$ to the optimizer, its real and imaginary parts of are concatenated into real valued $R^{2p \times p}$ matrix. The corresponding 
code is provided in the Supplementary Materials.

To apply ExFreDom to the $X \in R^{n \times p}$ data, first we take a Fourier transform of the data, 
then separate it into $M$ equal parts, where $M \in [5,10]$ and run (\ref{eq:fredomext}) minimization.

\section{Proof of Lemma~\ref{l:eqvar}} \label{a:proof}
Similar to \cite{chen2019}, we define $K = (k_{ji}) = (I - B)^{-1}$ such that the total effect
$k_{ji}$ is the sum over all possible directed paths from $i$ to $j$ of products of complex-valued coefficient
matrix $B$. From (\ref{eq:sden}), $\mbox{var}(\bol Y_j) = T^{-1} \sum_{l = 1}^{p} k_{jl} k^{*}_{lj}$.
Consequently, if $\Pi^G_j = \emptyset$ then $k_{jl} =0$ and $\mbox{var}(\bol Y_j) = T^{-1}$.
On the other hand, if $\Pi^G_j \neq \emptyset$ then there exist at least one parent  
$l \in \Pi^G_j$ whose descendants are not in the parent set, otherwise the acyclicity
assumption will be violated. For $l$, $k_{jl} k^{*}_{lj} = \beta_{jl} \beta^{*}_{jl} \geq \eta$
and the result follows.

\section{Additional Simulation Results} \label{a:addexp}
In this section we compare our FreDom algorithm with two recent time series structure learning algorithms:
  DYNOTEARS \citep{pamfil2020} and VARLINGAM \citep{hyvarinen2010}. In Appendix~\ref{a:tseqvar}, to show the advantage of the frequency domain
  approach over the time domain, we introduce TSEqVar algorithm, which similar to extensions
  proposed  in  \cite{hyvarinen2010,moneta2006,pamfil2020},
is a two-step time domain extension of the \cite{chen2019}.

The DYNOTEARS and VARLIGNAM algorithms are implemented using python packages \texttt{causalenex} and \texttt{lingam}, respectively. TSEqVar is a modification of
 the \texttt{EqVarDAG} R  package \citep{chen2019}. It is important to note that, DYNOTEARS and VARLIGNAM are
 designed for analyzing only real-valued time series. 

\begin{figure}
	\vskip 0.2in
	\centering
	\includegraphics[width = 0.5\columnwidth, height = 4cm]{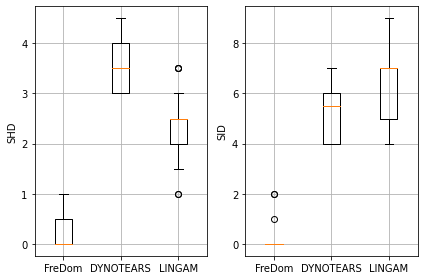}
	\caption{SHD and SID metrics for Experiment A.}
	 \label{f:exp2}
	 \vskip -0.2in
\end{figure}

\paragraph{Experiment A: Small Linear SVAR Model.}
In this experiment, we simulate data from the lag 1 linear SVAR $\Xt = B_0 \Xt + B_1\Xto + \ut$ model, where the 
adjacency matrices for $B_0$ and $B_1$ are 
$$
\begin{pmatrix}
	0& 1& 0& 0 &0\\
	0& 0& 0& 0& 0 \\
	0& 0& 0& 0& 0 \\
	0& 0& 1& 0& 0 \\
	0& 0& 0&1& 0\\
\end{pmatrix}
\mbox{ and }
\begin{pmatrix}
	0& 1& 1& 0 &0\\
	0& 0& 1& 0& 0 \\
	0& 0& 0& 0& 0 \\
	1& 0& 0& 0& 0 \\
	0& 0& 0& 1& 0\\
\end{pmatrix}
$$
Here $B_0$ entails
the instantaneous effects, and $u_i(t) \sim 0.4N(0,1)$. Figure~\ref{f:exp2} 
 reports the results. Here, FreDom performs the best.

\begin{figure}
 	\vskip 0.2in
	\centering
	\includegraphics[width = 0.5\columnwidth, height = 4cm]{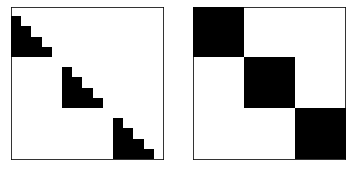}
	\caption{Non-zero pattern of instantaneous effect (Left) and coefficient matrix (Right) in Experiment B.}
	 \label{f:exp3pattern}
	 \vskip -0.2in
\end{figure}

\paragraph{Experiment B: Large linear SVAR Model.}
For $K = \{15,30\}$, we generate data from the SVAR($p$) model with $p = 3$ and 3 clusters (communities) of $K/3$ nodes each. Nodes 
within a community are not connected to any nodes in other communities. Figure~\ref{f:exp3pattern} illustrates the non-zero patterns 
for the instantaneous effects and block diagonal coefficient matrices, respectively. In Appendix~\ref{a:exp3detail}, we show that for such SVAR(p) process, 
when the covariance matrix of error terms is identity,
 the  inverse spectral density matrix has the same non-zero pattern as coefficient matrix (i.e.,  Figure~\ref{f:exp3pattern}(Right)
 and its Cholesky factor has the same non-zero pattern as the instantaneous effects. Consequently, from (\ref{eq:fcsem}), the true
 summary DAG is known.

 \begin{figure}
 	\vskip 0.2in
	\centering
	\includegraphics[width = 0.6\columnwidth, height = 8cm]{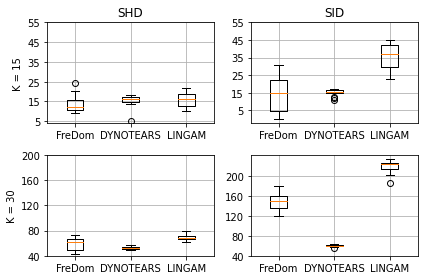}
	\caption{SHD and SID metrics for Experiment B. The rows correspond to K = \{15,30\}, respectively.}
	 \label{f:exp3}
	 \vskip -0.2in
\end{figure}

Figure~\ref{f:exp3} presents the results, in which rows corresponds to $K = \{15,30\}$ and columns to SHD and SID, respectively.
 We can see that for all $K$, FreDom reports the best SID and SHD metrics

\paragraph{Experiment C: Data from cSCM:}
In this experiment we generate complex-valued iid data from multiple linear cSCM that share the same DAG
structure. The true DAGs are simulated using the Erdos-Renyi model with the number of edges
equal to number of variables $p = 10$. The real and imaginary parts of the complex-valued coefficients are sampled from $[-2, -0.5] \cup [0.5,2]$ and the error term follows $N_c(0,1)$.

 \begin{figure}
 	\vskip 0.2in
	\centering
	\includegraphics[width = 0.6\columnwidth, height = 5cm]{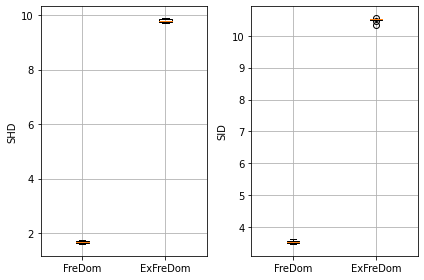}
	\caption{SHD and SID metrics for Experiment C.}
	 \label{f:exp3_c}
	 \vskip -0.2in
\end{figure}



\section{Details on Experiment B} \label{a:exp3detail}

In this section, we show that for a  SVAR(p) process generated as in Experiment B in Section~\ref{a:addexp} and with the identity
 covariance matrix of error terms, the inverse spectral density matrix and its Cholesky factor 
have the same non-zero pattern as the coefficient matrix and instantaneous effects, respectively.

 It is well known that for a VAR($p$) process with coefficient matrices $A_1, \dots, A_p$,
the inverse spectrum is a trigonometric matrix polynomial \citep{songsiri2009}

\begin{equation} \label{eq:cnd}
\Theta(\omega) =  \mathbf{A}^{H}(e^{i\omega}) \Sigma^{-1} \mathbf{A}(e^{i\omega}) = X_0 + \sum_{k = 1}^p(e^{-ik\omega}X_k + e^{ik\omega}X^{T}_k),
\end{equation}
 where $X_k = \sum_{i = 0}^{p - k}A^T_i \Theta A_{i + k}$ with $A_0 = I$.
Recalling that the relationship between VAR($p$) and SVAR($p$) coefficient matrices is $A_i = (I - B_0)^{-1}B_1$ \citep{hyvarinen2010} and plugging $\Sigma$ as an identity in (\ref{eq:cnd}), 
after some algebra the result follows. 

\section{Details on Experiment 1} \label{a:exp1}

The following proposition, which is a simple modification of \cite[Theorem 1]{dai2004}, states that (\ref{eq:ts}) can be used to generate a complex time series whose topological order and spectrum are identical to the given order and spectrum at Fourier frequencies.
\begin{proposition} \label{p:prop1}
 Let $S(\omega) = (1/T)(I_p - B(\omega) )^{-1}\{(I_p - B(\omega) )^{-1}\}^{H}$ be a positive definite spectral 
density matrix and the time series is generated as in (\ref{eq:ts}) then
\begin{enumerate}
	\item $S_T(\omega_k) = S(\omega_k)$ for $\omega_k = k /T$, where $S_T(\omega)$ is the spectrum of $\Xt$
	\item if $S(\omega)$ has continuous second derivative with respect to $\omega$, then, for any $\omega \in [0,1]$,
	$|S_T(\omega) - S(\omega)| = O(T^{-1})$
\end{enumerate}
 \end{proposition}


\section{Air Pollution Additional Results} \label{a:airadd}
\begin{figure}[!ht]
  \vskip 0.2in
   \begin{subfigure}[b]{0.25\textwidth}%
      \includegraphics[width=\textwidth,height = 2.8cm]{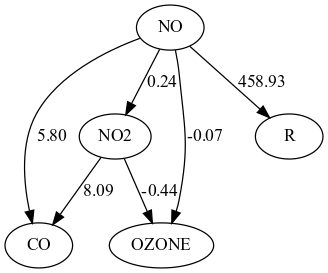}
    \end{subfigure}
    \begin{subfigure}[b]{0.25\textwidth}%
      \includegraphics[width=\textwidth,height = 2cm]{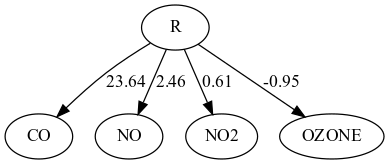}
    \end{subfigure}
    \begin{subfigure}[b]{0.25\textwidth}%
      \includegraphics[width=\textwidth,height = 2.5cm]{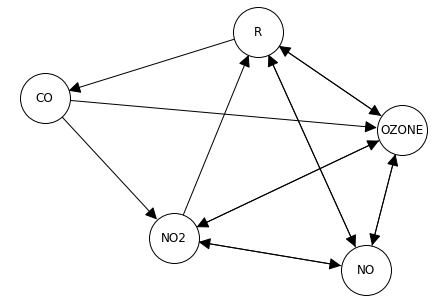}
    \end{subfigure}
    \caption{Additional results for the air pollution data. (a) LINGAM (b) NOTEARS
    (c) Granger} \label{f:sumdag_a}
    \vskip -0.2in
  \end{figure}

\begin{figure}
	\vskip 0.2in
	\centering
	\includegraphics[width = 0.6\columnwidth, height = 6.5cm]{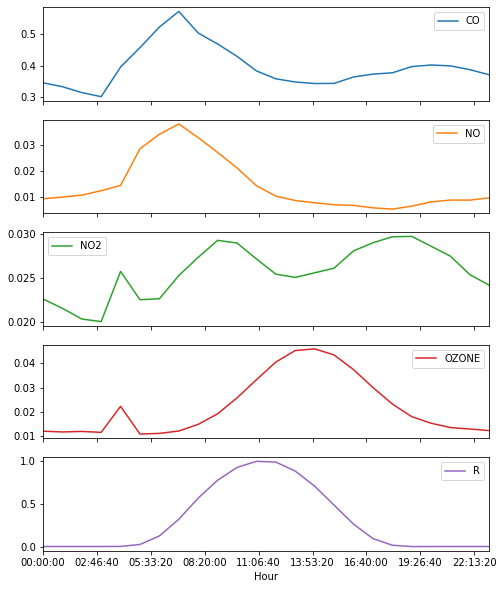}
	\caption{\titlecap{Average of daily measurements for the five pollutants}}
	 \label{f:aveplot}
	 \vskip -0.2in
\end{figure}

In this section we report an additional results for the Air Pollution dataset
discussed in Section~\ref{s:air}.
Figure~\ref{f:aveplot} reports the average of daily measurements and in Figure~\ref{f:sumdag_a} we  report the instantaneous effects obtained from the LINGAM, DYNOTEARS, and the Granger Causality Graph. The weights on the edges in  Figure~\ref{f:sumdag_a}(a,b) show the corresponding estimated coefficients of  instantaneous effects.  

 \section{Stock Return Volatility Data} \label{a:stock}
\begin{figure}
	\vskip 0.2in
	\centering
	\includegraphics[width = 0.55\columnwidth, height = 6cm]{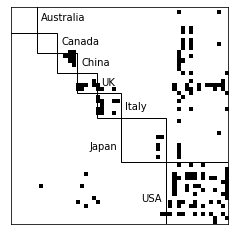}
	\caption{Estimated adjacency matrix with rows and columns sorted by country.}
	 \label{f:stocknotdag}
	 \vskip -0.2in
\end{figure}

Figure~\ref{f:stocknotdag} illustrate the summary graph for the ExFreDom algorithm.

\section{TSEqVar algorithm} \label{a:tseqvar}

TSEqVar is a two-step, time domain extension of the EqVar \citep{chen2019}.  
Recall that linear SVAR with lag $p_0$ is given 
\begin{equation} \label{eq:linsvar11}
	\Xt = B_0\Xt  + B_1 \Xto + \dots B_{p_0}\Xtp +  \bol\varepsilon(t),
\end{equation}
and the relation between VAR and SVAR coefficient matrices is $B_i = (I - B_0)A_i$.
Similar to \cite{hyvarinen2010,moneta2006,pamfil2020}, TSEqVar implements the following two steps:
\begin{enumerate}
\item Run VAR($p_0$) model to obtain residuals and $\hat A_i$ 
\item Run EqVar algorithm on residuals to recover matrix $\hat B_0$ and $\hat B_i = (I - \hat B_0) \hat A_i$
\end{enumerate}

Table~\ref{t:tseqvar} report the simulation results for TSEqVar for Experiments 2 and 4.

\begin{table}[ht!]
\centering
	 \caption{Mean and Standard Deviations for Experiments B and C  over 50 repetitions.}
\begin{tabular}{c|c|c|c}
	\hline
	&K&SHD $\tau$ & SID \\
	\hline
	\multirow{2}{*}{Exp. B}&15 &32.33(13.52) &	28.33(16.32)\\
				 	   &30&  135.80(35.36)& 	125.95(34.89)\\
	\hline
	\multirow{1}{*}{Exp. C}&4 &6.05(1.22) &	8.95(3.36)\\
	\hline
	 \end{tabular}
	 	 \label{t:tseqvar}
\end{table}

\end{document}